\documentclass[10pt,twoside,twocolumn,english,aps,prl,superscriptaddress,notitlepage]{revtex4-1}
\usepackage{mathpazo}
\usepackage{helvet}

\usepackage[LGR,T1]{fontenc}
\usepackage[latin9]{inputenc}
\usepackage[a4paper]{geometry}
\usepackage{xcolor}
\usepackage{pdfcolmk}
\usepackage{babel}
\usepackage{verbatim}
\usepackage{textcomp}
\usepackage{amssymb}
\usepackage{graphicx}
\PassOptionsToPackage{normalem}{ulem}
\usepackage{ulem}
\usepackage[unicode=true,pdfusetitle,
 bookmarks=false,
 breaklinks=true,pdfborder={0 0 0},pdfborderstyle={},backref=false,colorlinks=true]
 {hyperref}
\hypersetup{
 colorlinks=true,citecolor=blue,linkcolor=blue,urlcolor=blue,pagebackref=true}

\makeatletter

\DeclareRobustCommand{\greektext}{%
  \fontencoding{LGR}\selectfont\def\encodingdefault{LGR}}
\DeclareRobustCommand{\textgreek}[1]{\leavevmode{\greektext #1}}
\ProvideTextCommand{\~}{LGR}[1]{\char126#1}

\providecolor{lyxadded}{rgb}{0,0,1}
\providecolor{lyxdeleted}{rgb}{1,0,0}
\DeclareRobustCommand{\lyxsout}[1]{\ifx\\#1\else\sout{#1}\fi}

\usepackage{amsfonts,xcolor,graphicx,colortbl}
\usepackage[calcwidth,explicit]{titlesec}\usepackage[normalem]{ulem}
\makeatletter\renewcommand\frontmatter@abstractwidth{\dimexpr\textwidth-2cm\relax}\makeatother

\renewcommand\thesection{\Alph{section}}%\renewcommand\thesubsection{\Alph{section}\Arabic{subsection}}

\makeatletter\@addtoreset{paragraph}{section}\makeatother
\makeatletter\def\p@paragraph{}\makeatother

\AtBeginDocument{
\renewcommand{\ref}[1]{\autoref{#1}}

}

\makeatletter\def\l@section{\@dottedtocline{1}{0.6em}{1.5em}}\makeatother
\makeatletter\def\l@paragraph{\@dottedtocline{4}{1.5em}{1.8em}}\makeatother
\makeatletter\def\l@figure{\@dottedtocline{1}{0.6em}{1.8em}}\makeatother

\titleformat{\section}{\normalfont\bfseries\scshape\filcenter}{\thesection.}{0.25em}{#1}[{\titlerule[0.2pt]}]
\titlespacing*{\section}{0pt}{0.25ex}{0.25ex}
\titleformat{\subsection}{\bfseries\sffamily}{\thessubsection.}{1em}{#1}
\titlespacing{\subsection}{0pt}{1ex}{0.5ex}
\titleformat{\paragraph}[runin]{\normalsize\sffamily\bfseries}{}{0pt}{}
\titlespacing*{\paragraph}{1em}{-1ex}{0em}[] %11pt: 1.25em, 10pt: 1em

\addto\captionsenglish{}

\setcitestyle{super}%\setlength{\bibsep}{0.1mm}\setlength{\bibhang}{0pt}

\makeatother

\begin{document}

\title{Sub-100 nm Skyrmions at Zero Magnetic Field in Ir/Fe/Co/Pt Nanostructures
\smallskip{}
}

\author{Pin Ho}

\affiliation{Data Storage Institute, 2 Fusionopolis Way, 138634 Singapore}

\author{Anthony K.C. Tan}

\affiliation{Data Storage Institute, 2 Fusionopolis Way, 138634 Singapore}

\affiliation{Division of Physics and Applied Physics, School of Physical and Mathematical
Sciences, Nanyang Technological University, 637371 Singapore}

\author{S. Goolaup}

\affiliation{Data Storage Institute, 2 Fusionopolis Way, 138634 Singapore}

\author{A.L. Gonzalez Oyarce}

\affiliation{Data Storage Institute, 2 Fusionopolis Way, 138634 Singapore}

\author{M. Raju}

\affiliation{Division of Physics and Applied Physics, School of Physical and Mathematical
Sciences, Nanyang Technological University, 637371 Singapore}

\author{L.S. Huang}

\affiliation{Data Storage Institute, 2 Fusionopolis Way, 138634 Singapore}

\author{Anjan Soumyanarayanan}

\thanks{Correspondence should be addressed to \textbf{A.S. }(souma@dsi.a-star.edu.sg)
or \textbf{C.P. }(christos@ntu.edu.sg)}

\affiliation{Data Storage Institute, 2 Fusionopolis Way, 138634 Singapore}

\affiliation{Division of Physics and Applied Physics, School of Physical and Mathematical
Sciences, Nanyang Technological University, 637371 Singapore}

\author{C. Panagopoulos}

\thanks{Correspondence should be addressed to \textbf{A.S. }(souma@dsi.a-star.edu.sg)
or \textbf{C.P. }(christos@ntu.edu.sg)}

\affiliation{Division of Physics and Applied Physics, School of Physical and Mathematical
Sciences, Nanyang Technological University, 637371 Singapore}
\begin{abstract}
\noindent Magnetic skyrmions are chiral spin structures that have
recently been observed at room temperature (RT) in multilayer thin
films. Their topological stability should enable high scalability
in confined geometries \textendash{} a sought-after attribute for
device applications. While umpteen theoretical predictions have been
made regarding the phenomenology of sub-100 nm skyrmions confined
in dots, in practice their formation in the absence of an external
magnetic field and evolution with confinement remain to be established.
Here we demonstrate the confinement-induced stabilization of sub-100
nm RT skyrmions at zero field (ZF) in Ir/Fe$(x)$/Co$(y)$/Pt nanodots
over a wide range of magnetic and geometric parameters. The ZF skyrmion
size can be as small as \textasciitilde{}50 nm, and varies by a factor
of 4 with dot size and magnetic parameters. Crucially, skyrmions with
varying thermodynamic stability exhibit markedly different confinement
phenomenologies. These results establish a comprehensive foundation
for skyrmion phenomenology in nanostructures, and provide immediate
directions for exploiting their properties in nanoscale devices.
\end{abstract}
\maketitle
\begin{comment}
\textbf{\uline{Buffer Text}}
\begin{itemize}
\item This is imperative for integration into technologically useful applications. 
\end{itemize}
\end{comment}

\noindent 
\section{Introduction}

\noindent %
\begin{comment}
\textbf{A1. Sk Properties:}
\begin{itemize}
\item A
\end{itemize}
\end{comment}

\paragraph{Sk Properties}

\noindent The whirling arrangement of spins that defines a magnetic
skyrmion arises from chiral relativistic interactions in magnetic
materials lacking inversion symmetry\citep{Bogdanov2001,Nagaosa2013}.
The topologically protected spin structure of skyrmions manifests
in their emergent behavior as distinct magnetic quasiparticles\citep{Soumyanarayanan2016},
with individual addressability, and current-induced creation and dynamics\citep{Romming2013,Sampaio2013}.
The discovery of RT skyrmions in multilayer films\citep{MoreauLuchaire2016,Woo2016,Boulle2016,Jiang2015,Soumyanarayanan2017},
material platforms of demonstrable technological relevance, has generated
an explosion of interest in investigating these topological quasiparticles
in device-relevant configurations\citep{Kang2016,Finocchio2016,Fert2017}.
In particular, device proposals build upon their lateral mobility
in wires\citep{Sampaio2013,Iwasaki2013,Zhang2015h}, and their manipulation
in dots\citep{Sampaio2013,Finocchio2015,Zhang2015}. Harnessing the
potential of nanoscale skyrmions calls for an immediate understanding
of their phenomenology in confined geometries. %
\begin{comment}
\textbf{\uline{A2. Sk-Dots: Motivation \& Previous Work:}}
\begin{itemize}
\item A
\end{itemize}
\end{comment}

\paragraph{Motivation \& Previous Work}

The interest surrounding skyrmions confined in dots is piqued by the
knowledge amassed on conventional nanomagnetic dots, and their technological
relevance\citep{Wolf2010}. The topological stability of skyrmions\citep{Hagemeister2015}
could enable dot devices with nanometer scalability\citep{Romming2013},
with ease of detection\citep{Romming2013,Hanneken2015,Tomasello2017a},
and energy-efficient manipulation\citep{Romming2013,Zhang2015,Nakatani2016,Bhattacharya2017}.
In particular, they could be used in magnetic tunnel junction (MTJ)-like
configurations, with applications in memory\citep{Nakatani2016,Bhattacharya2017},
logic\citep{Kang2016}, oscillators\citep{Zhou2015,Garcia-Sanchez2016},
and microwave detectors\citep{Finocchio2015}. This has prompted a
flurry of theoretical and experimental efforts to establish the phenomenology
of confined skyrmions. Notably, theoretical works predict the scaling
of skyrmion size and stability with geometric and magnetic parameters\citep{Rohart2013,Sampaio2013,Guslienko2015,Mulkers2017,Kolesnikov2017,Tomasello2017,Zelent2017}.
Experimental studies on multilayers have reported $>150$~nm skyrmion
bubbles in the absence of external fields, with the chiral spin structure
stabilized by the additional presence of dipolar interactions at larger
length scales\citep{Boulle2016,Woo2016,Zeissler2017}. However, skyrmions
with sub-100~nm sizes, thus far stabilized at finite external magnetic
fields\citep{MoreauLuchaire2016,Soumyanarayanan2017}, have yet to
exhibit confinement effects. Till date, the zero field (ZF) stabilization
of sub-100~nm skyrmions in dots, the role of confinement in governing
skyrmion properties, and its interplay with magnetic parameters \textendash{}
remain to be established. %
\begin{comment}
\textbf{A4. Summary of Results}
\begin{itemize}
\item A
\end{itemize}
\end{comment}

\paragraph{Summary of Results}

Here we report the ZF stabilization of sub-100~nm RT skyrmions in
nanodots of Ir/Fe$(x)$/Co$(y)$/Pt multilayer films. We utilize magnetic
force microscopy (MFM) and micromagnetic simulations to establish
their stability over a wide range of magnetic and geometric parameters.
In our patterned dots, the ZF skyrmion size, as small as $\sim50$~nm,
varies by a factor of $4$ with confinement and magnetic interactions.
In particular, confined skyrmions exhibit markedly different phenomenologies
with varying thermodynamic stability. Our results provide immediate
directions for tailoring skyrmions properties to confined geometries
geared to device applications.

\noindent 
\section{Confined States in Multilayers\label{sec:SkConfinement}}

\noindent %
\begin{comment}
\textbf{\uline{B1. Skyrmion Energetics: Buffer Text}}
\begin{itemize}
\item A
\end{itemize}
\end{comment}
\begin{figure}
\begin{centering}
\includegraphics[width=3.2in]{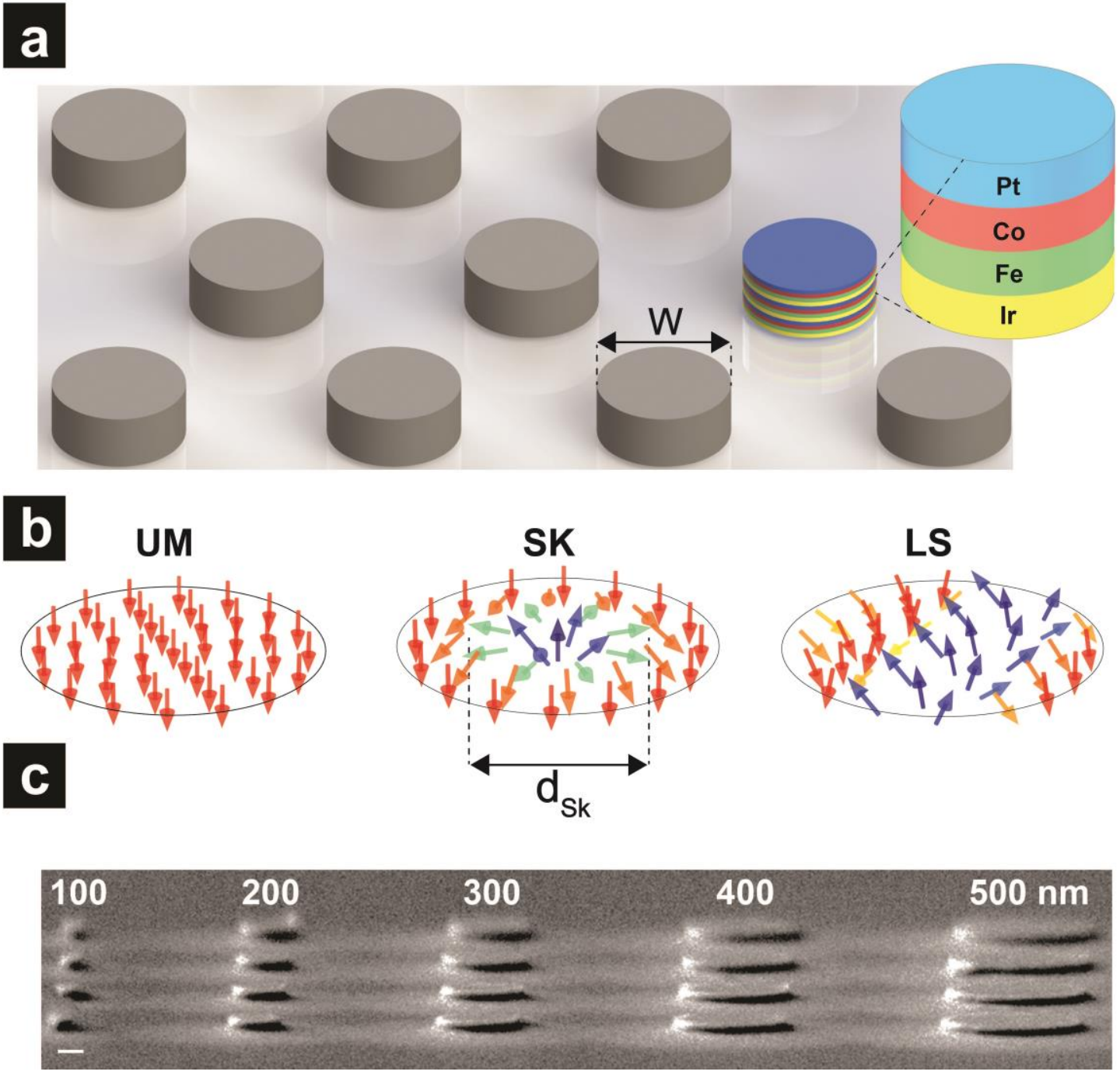}
\par\end{centering}
\noindent \caption[Schematic Magnetic States in Dots]{\textbf{Confined Magnetic States in Multilayer Nanodots. (a)} Schematic
array of sub-\textgreek{m}m dots (diameter $W$) composed of Ir/Fe$(x)$/Co$(y)$/Pt
multilayers. Inset shows the sequence of Ir, Fe, Co, and Pt layers
forming a stack.\textbf{ (b)} Schematic spin textures corresponding
to the three distinct states expected in such nanodots: uniform magnetization
(UM: left), Néel skyrmion with size $d_{{\rm Sk}}$ (SK: centre),
and labyrinthine stripes (LS: right). \textbf{(c)} Cross-sectional
scanning electron microscope (SEM) image (scale bar: 100~nm) of a
{[}Ir/Fe/Co/Pt{]}$_{20}$ dot array with $W$ from 100 \textendash{}
500 nm.\label{fig:SkDot_Schematic}}
\end{figure}

\paragraph{Confined Skyrmion Energetics}

\noindent The ground state configuration of a magnetic multilayer
nanostructure is determined by the confluence of magnetic interactions.
The exchange interaction, characterized by the stiffness ($A$), aligns
neighboring spins parallel, and favors a uniformly magnetized (UM)
state (\ref{fig:SkDot_Schematic}b, left), with orientation determined
by the effective out-of-plane (OP) anisotropy, $K_{{\rm eff}}$. In
contrast, the interfacial Dzyaloshinskii-Moriya interaction (DMI,
$D$) prefers a winding spin arrangement, leading to a labyrinthine
stripe (LS) state (\ref{fig:SkDot_Schematic}b, right)\citep{Heide2008}.
The competition between $D$, $A$, and $K_{{\rm eff}}$ can form
Néel-textured skyrmions (SK, \ref{fig:SkDot_Schematic}b, centre)\citep{Romming2013},
thermodynamically stable entities for $\kappa=\pi D/4\sqrt{AK_{{\rm eff}}}>1$\citep{Bogdanov2001,Soumyanarayanan2016}.
Notably, confined geometries can dramatically influence the ground
state configuration\citep{Rohart2013,Beg2015,Zhao2016,Boulle2016,Jin2017,Vidal-Silva2017},
potentially favoring the ZF stabilization of skyrmions\citep{Rohart2013,Sampaio2013}.
The added presence of interlayer dipolar coupling in multilayer stacks\citep{Vidal-Silva2017,Pulecio2016,Zelent2017},
and long-range intralayer dipolar interactions\citep{Boulle2016},
can also influence the ground state \textendash{} the latter being
key to the pioneering works reporting the stability of confined skyrmion
'bubbles'\citep{Jiang2015,Boulle2016,Woo2016,Zeissler2017,Zelent2017}.
Motivated by this, we focus on the ZF stabilization of sub-100~nm
Néel skyrmions in confined geometries\citep{Fert2017}. %
\begin{comment}
\textbf{\uline{B2. Multilayer Platform for Controlling Skyrmions:
Buffer Text}}
\begin{itemize}
\item A
\end{itemize}
\end{comment}
\begin{figure}[h]
\begin{centering}
\includegraphics[width=3.2in]{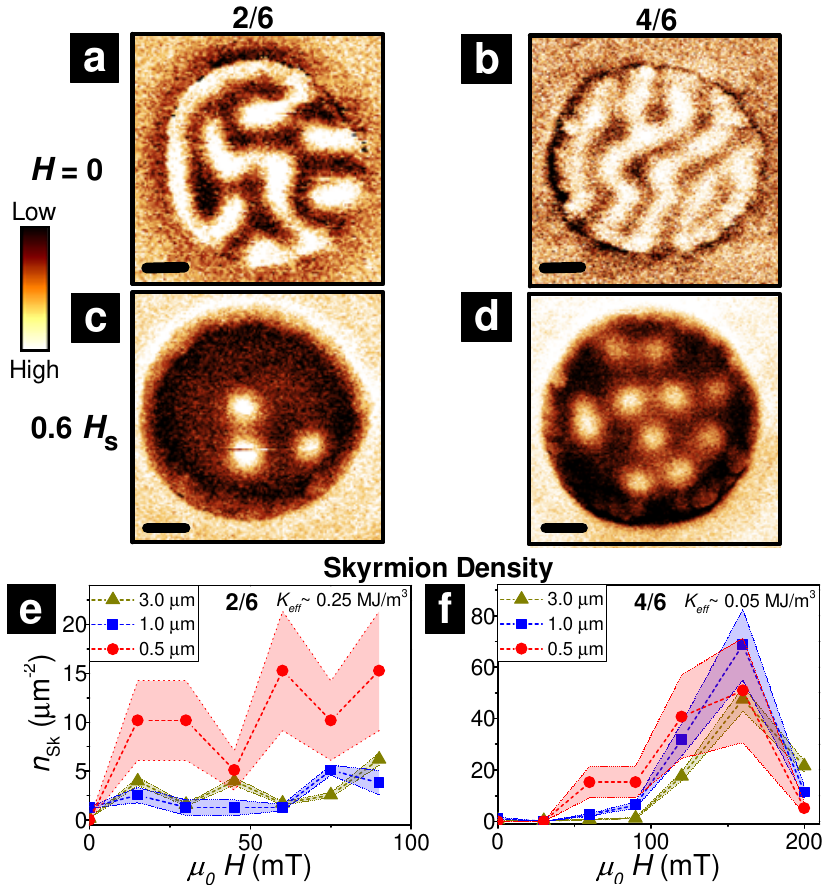}
\par\end{centering}
\noindent \caption[Confinement Effects in Sub-\textgreek{m}m Dots]{\textbf{Confinement Effects in Sub-\textgreek{m}m Dots. (a-d) }MFM
images (scale bar: 100~nm) of $W=500$~nm dots after negative saturation
for Fe(2)/Co(6) (a, c) and Fe(4)/Co(6) (b, d) stacks, respectively.
At $H=0$, (a-b) show LS states; At $H\simeq0.6\,H_{S}$ (c-d) show
SK states, in isolated (c) and lattice (d) configurations, respectively.\textbf{
(e-f) }Measured skyrmion density, $n_{{\rm Sk}}$, as a function of
$H$, for dots with varying $W$ for samples Fe(2)/Co(6) (e) and Fe(4)/Co(6)
(f) respectively. Shaded regions represent error bars.\label{fig:SkDots_Density}}
\end{figure}

\paragraph{Materials \& Methods}

The stability and size of confined skyrmions is expected to be markedly
influenced by magnetic interactions (e.g. $D$, $A$, and $K_{{\rm eff}}$)
and geometric parameters (here, $W$)\citep{Rohart2013,Zelent2017,Vidal-Silva2017}.
The multilayer Ir/Fe$(x)$/Co$(y)$/Pt, wherein magnetic interactions
can be tailored by the Fe$(x)$/Co$(y)$ composition, is an aptly
suited platform for establishing confined skyrmion phenomenology.
{[}Ir$(10)$/Fe$(x)$/Co$(y)$/Pt$(10)${]}$_{20}$ stacks (layer
thicknesses in Å in parentheses) host Néel-textured skyrmions\citep{Yagil2017}
with tunable size ($2\times$), thermodynamic stability ($10\times$),
and density ($10\times$) across compositions\citep{Soumyanarayanan2017}
(details in \ref{sec:Methods}). Here we investigate confined skyrmions
in dots ($W$: 100 \textendash{} 3000~nm) patterned from these stacks
(e.g. \ref{fig:SkDot_Schematic}c, details in \ref{sec:Methods}).
The five samples studied here, described henceforth by their Fe($x$)/Co($y$)
composition, allow us to vary $D/A$ from 15 \textendash{} 18~nm$^{-1}$
and $K_{{\rm eff}}$ from 0.01 \textendash{} 0.25~MJ/m$^{3}$, corresponding
to an order of magnitude modulation of $\kappa$ (0.8\textendash 6.3,
\ref{fig:ZFSkyrmions_Expts}a, details in \textcolor{blue}{§SI 1}).
The magnetic configuration was imaged by MFM using ultra-low moment
tips for high-resolution, non-perturbative imaging, after negative
OP saturation (see \ref{sec:Methods}). The excellent agreement observed
between the magnetic texture evolution for our $W\geq1000$~nm dots
with corresponding film-level results (\textcolor{blue}{§SI 3}) establishes
a firm foundation for investigating confinement effects in sub-\textgreek{m}m
geometries.%
\begin{comment}
\textbf{\uline{B3. Confinement \& Sk Density Buffer Text}}

\noindent A
\end{comment}

\paragraph{Sk Density in Sub-\textgreek{m}m Dots}

\noindent We begin by examining the emergence of confinement effects
as $W$ is reduced to 500~nm. Consistent with film-level results\citep{Soumyanarayanan2017},
we observe an LS state at ZF (\ref{fig:SkDots_Density}a,b) transforming
into sub-100~nm skyrmions at finite OP fields ($H$, \ref{fig:SkDots_Density}c,d).
However, the skyrmion density, $n_{{\rm Sk}}$, displays a striking
contrast with $W$-evolution across samples. For Fe(2)/Co(6) ($K_{{\rm eff}}=0.25$~mJ/m$^{3}$,
\ref{fig:SkDots_Density}e), $n_{{\rm Sk}}$ is consistently higher
in 500~nm dots, by a factor of 3, as compared to larger dots and
films. This increase in $n_{{\rm Sk}}$ with smaller $W$ can be attributed
to the enhancement of the in-plane (IP) demagnetization field with
confinement\citep{Woo2016,Boulle2016}. Meanwhile, for Fe(4)/Co(6)
($K_{{\rm eff}}=0.06$~mJ/m$^{3}$, \ref{fig:SkDots_Density}f),
$n_{{\rm Sk}}(H)$ is consistent across $W$ down to $500$~nm, and
$\sim5-10\times$ higher than for Fe(2)/Co(6). Indeed $n_{{\rm Sk}}$
is known to increase with reducing $K_{{\rm eff}}$ as the energy
barrier for domain nucleation is lowered\citep{Soumyanarayanan2017}.
Importantly, the constancy of $n_{{\rm Sk}}$ with $W$ for Fe(4)/Co(6)
\textendash{} which already hosts a dense skyrmion lattice \textendash{}
offers an orthogonal tuning parameter. This suggests that utilizing
the synergy between magnetic tuning and confinement in sub-500~nm
dots is a promising route for ZF skyrmion stabilization.

\noindent \noindent 
\begin{figure*}[t]
\begin{centering}
\includegraphics[width=6.6in]{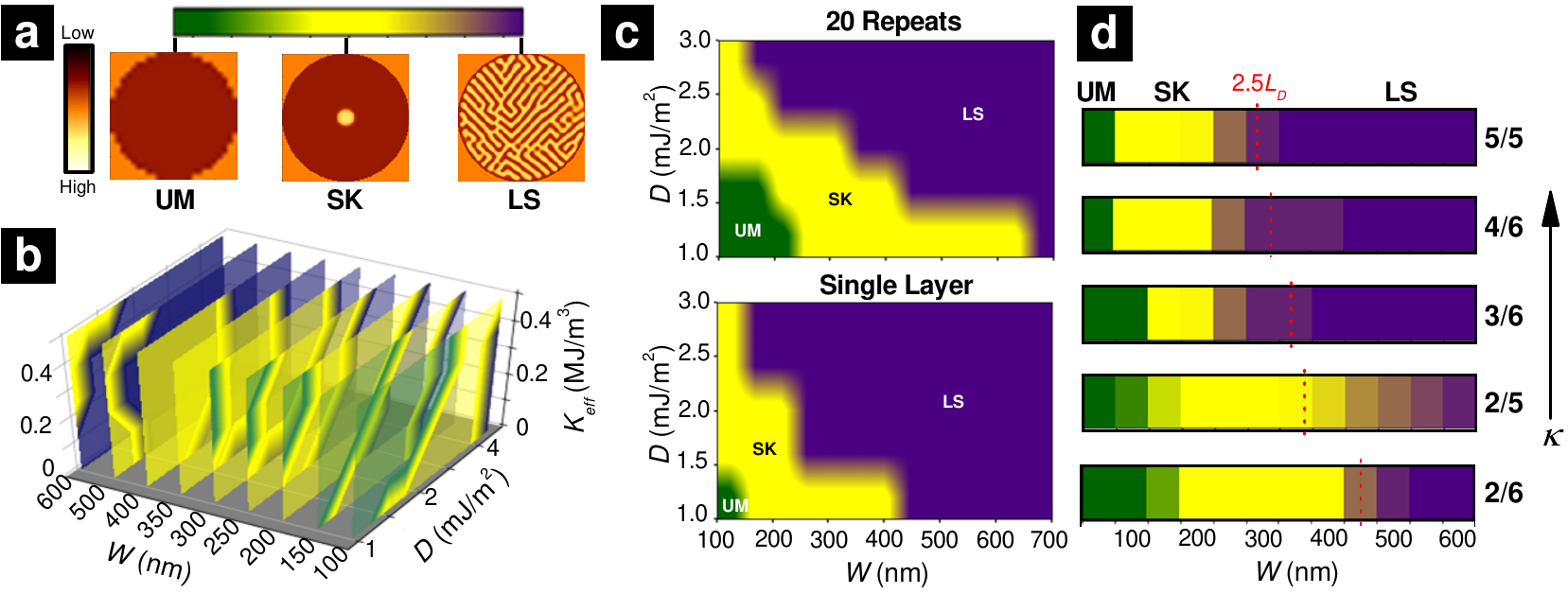}
\par\end{centering}
\noindent \caption[Simulated Evolution of ZF Magnetic States]{\textbf{Simulated Evolution of ZF Magnetic States. }Micromagnetic
simulations of the ZF phase diagram for {[}Ir/Fe$(x)$/Co$(y)$/Pt{]}$_{20}$
dots. \textbf{(a)} Magnetic textures obtained (examples shown) are
identified as UM (left: green), SK (centre: yellow), or LS (right:
blue) respectively. Phase diagrams with Fe(2)/Co(6) parameters, for:
\textbf{(b)} 20 repeats, where $D$, $K_{{\rm eff}}$, and $W$ are
varied over a range of likely values; \textbf{(c)} 20 repeats (top)
and single layer (bottom): $D$ and $W$ are varied over a range of
likely values. SK is the expected ground state over a broad, intermediate
set of parameters. \textbf{(d)} Expected evolution of magnetic states
in dots with $W$ ranging from 50 \textendash{} 600~nm across the
five samples (shown with increasing $\kappa$, parameters in \ref{fig:ZFSkyrmions_Expts}a,
\textcolor{blue}{§SI 1}).\label{fig:ZFStates_MuMag}}
\end{figure*}

\section{Confined Zero Field Skyrmions\label{sec:ZFSkyrmions}}

\noindent %
\begin{comment}
\textbf{\uline{C1. ZF Skyrmion Simulations: Buffer Text}}
\begin{itemize}
\item A
\end{itemize}
\end{comment}

\paragraph{Magnetic Phase Diagram}

\noindent We performed a comprehensive set of multilayer micromagnetic
simulations to map the evolution of magnetic states for Ir/Fe$(x)$/Co($y$)/Pt
dots (details in \ref{sec:Methods}, parameters in \textcolor{blue}{§SI
1}). The relaxed magnetic state at ZF, following the introduction
of a skyrmion, was examined over a range of parameters to determine
the magnetic phase diagram\citep{Sampaio2013,Woo2016}. \ref{fig:ZFStates_MuMag}b
shows such a ZF phase diagram (magnetic states in \ref{fig:ZFStates_MuMag}a)
with varying $W$ (100-600~nm), $D$ ($1-4$~mJ/m$^{2}$), $K_{{\rm eff}}$
($0-0.5$~MJ/m$^{3}$), and nominal parameters of Fe(2)/Co(6) (\textcolor{blue}{§SI
4}). While the LS phase is observed at large $W$, the reduction in
magnetostatic energy for $W\lesssim500$~nm shrinks the stable domain
wall size, and instead favors the formation of an SK phase\citep{Rohart2013}.
As $W$ is reduced further (e.g. below $\sim100$~nm), the exchange
energy eventually dominates, leading to the UM phase. Moreover, the
interplay between confinement and magnetic parameters determines the
window for SK stability in dots, which is explored here.%
\begin{comment}
\textbf{\uline{C2. MH \& SkProps: Buffer Text}}
\begin{itemize}
\item A
\end{itemize}
\end{comment}

\paragraph{Sk Stability: Simulation Trends}

First, a comparison of the phase diagram for the Fe(2)/Co(6) multilayer
(\ref{fig:ZFStates_MuMag}c, top) with the corresponding single layer
(\ref{fig:ZFStates_MuMag}c, bottom) shows the SK phase persisting
over a much larger $W$ range in the multilayer. The addition of interlayer
dipolar coupling, introduced by multilayer stacking, is key to this
increased SK stability\citep{Pulecio2016,Zelent2017}. Next, an inspection
of the confined magnetic states across samples (\ref{fig:ZFStates_MuMag}d,
optimal parameters) suggests that ZF skyrmions should be observable
for all compositions. Finally, the SK phase may be observed for $W\lesssim2.5\,L_{{\rm D}}$
($L_{{\rm D}}$ is the film-level domain periodicity), and its stability
in smaller dots could be enhanced with increasing $\kappa$. These
results provide promise and specific directions for stabilizing skyrmions
in Ir/Fe$(x)$/Co$(y)$/Pt nanostructures. 
\begin{figure*}
\begin{centering}
\includegraphics[width=6.6in]{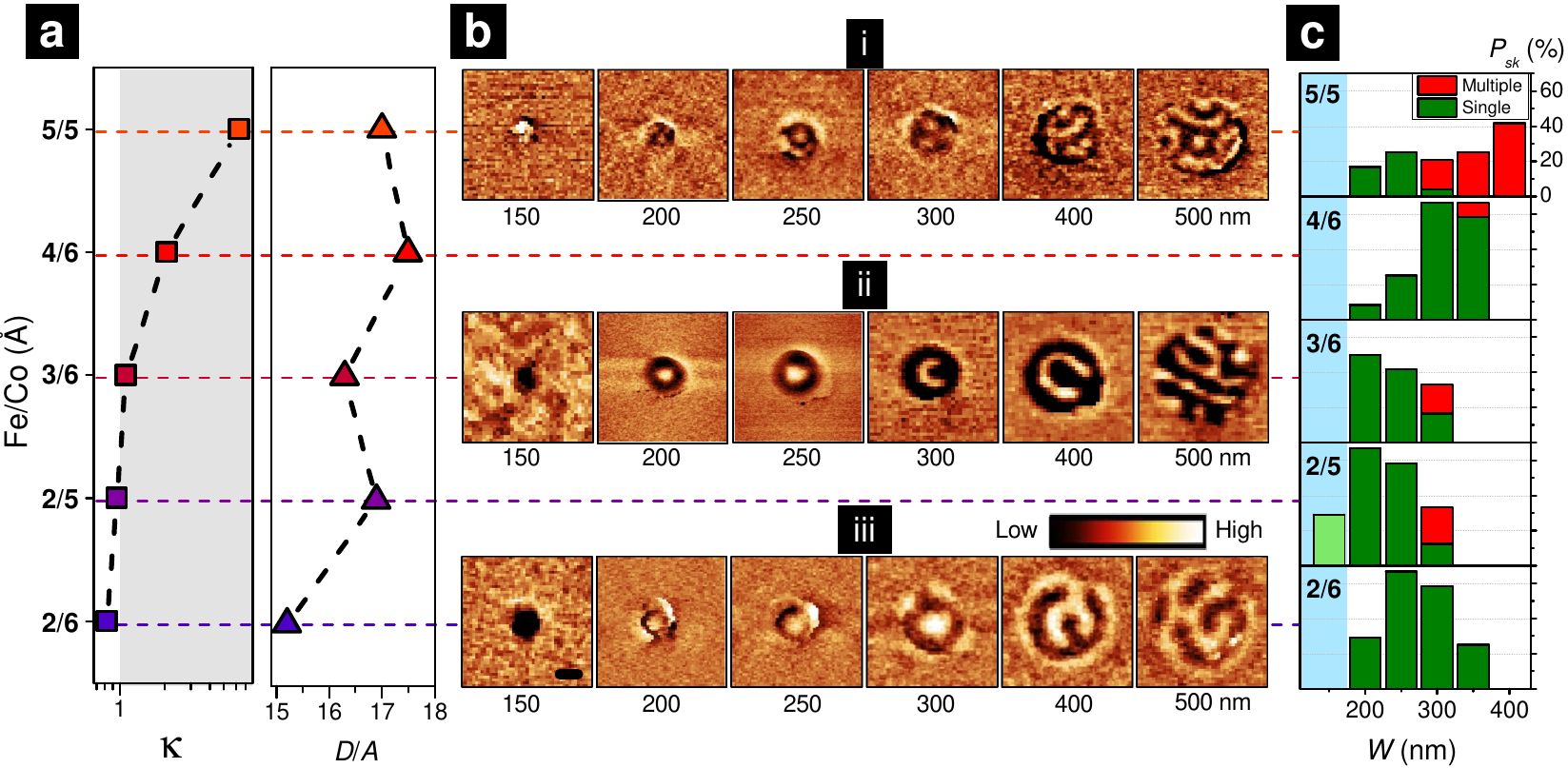}
\par\end{centering}
\noindent \caption[Imaging Confined ZF Skyrmions]{\textbf{Imaging Confined ZF Skyrmions.} \textbf{(a)} Variation of
the parameters: $\kappa$ and $D/A$ across the studied samples. \textbf{(b)}
MFM images at ZF (scale bar: 100~nm) showing the magnetic states
in dots with $W$ ranging from 150 \textendash{} 500 nm, for samples
Fe(2)/Co(6), Fe(3)/Co(6), and Fe(5)/Co(5) respectively. The images,
in several cases, show ZF skyrmions at intermediate $W$, while UM
and LS phases are observed for smaller and larger dots respectively.
\textbf{(c)} Empirical ZF skyrmion nucleation probability, $P_{{\rm Sk}}$
(averaged over 12 nominally identical dots), as a function of $W$
for all five samples. Shaded blue region ($W<200$~nm) indicates
a low yield from the patterning process. Both single (green bars)
and multiple (red bars) skyrmion configurations are observed.\label{fig:ZFSkyrmions_Expts}}
\end{figure*}
\begin{comment}
\textbf{\uline{C3. Mag-Sk Props - T-Dep: Buffer Text}}
\begin{itemize}
\item A
\end{itemize}
\end{comment}

\paragraph{Confined ZF Skyrmions}

We now turn to MFM images of ZF magnetic textures for $W<500$~nm
dots, shown for three samples in \ref{fig:ZFSkyrmions_Expts}b (remainder
in \textcolor{blue}{§SI 3}). Consistent with simulations (\ref{fig:ZFStates_MuMag}d),
reducing $W$ ($500-150$~nm: \ref{fig:ZFSkyrmions_Expts}b, right
to left), results in a gradual transition from LS to SK phases, and
eventually to the UM phase. Crucially, sub-100~nm skyrmions are stabilized
at ZF, \emph{prima facie} by confinement effects, across all Ir/Fe$(x)$/Co$(y)$/Pt
compositions. In some cases, however, nominally identical dots are
found to exhibit different ZF states, likely due to the granularity
of sputtered multilayer films\citep{Legrand2017,Zeissler2017,Juge2017},
or fabrication process variations (\ref{sec:Methods}, \textcolor{blue}{§SI
2}). This variability is mitigated by determining the statistically
averaged behavior of 12 dots for each $W$ across samples. The evolution
of ZF skyrmion state probability thus obtained, $P_{{\rm Sk}}$, is
further examined. %
\begin{comment}
\textbf{\uline{C4. ZF Nucleation: Buffer Text}}
\begin{itemize}
\item A
\end{itemize}
\end{comment}

\paragraph{Sk Stability: Expt Trends}

Histogram plots of $P_{{\rm Sk}}(W)$ across samples (\ref{fig:ZFSkyrmions_Expts}c)
evidence a stable SK phase over a wide range of magnetic (\ref{fig:ZFSkyrmions_Expts}a)
and geometric parameters. In line with simulations (\ref{fig:ZFStates_MuMag}d),
the increased skyrmion stability in multilayers underscores the vital
role of interlayer dipolar interactions. Next, while the peak $P_{{\rm Sk}}$
appears to shift to lower $W$ with $\kappa$ for $\kappa\lesssim1$,
this trend, expected from simulations, does not persist for $\kappa>1$.
For $W<200$~nm (\ref{fig:ZFSkyrmions_Expts}c shaded region), low
yield in our patterning process precludes a statistically meaningful
comparison across $\kappa$ (\ref{sec:Methods}). More surprising
is the persistence of the SK phase for $\kappa>1$ at larger $W$,
and particularly, the observation of multi-skyrmion (m-SK) configurations
(e.g. \ref{fig:ZFSkyrmions_Expts}b-i, 300 nm; \ref{fig:ZFSkyrmions_Expts}c,
red). While thermodynamically stable skyrmions ($\kappa>1$) form
ordered lattices at finite fields\citep{Bogdanov2001,Romming2013,Nagaosa2013,Zhao2016,Soumyanarayanan2017},
the emergence of m-SK configurations at ZF emphasizes a strong interplay
of magnetic and confinement effects that was not considered in previous
simulations. Indeed, when simulations are repeated with the initialized
skyrmion number, $N_{{\rm i}}>1$ (\ref{fig:SkSize_Sims}d, \textcolor{blue}{§SI
4})\citep{Beg2015,Woo2016} \textendash{} m-SK states are found to
be stable for $\kappa>1$ dots (\ref{fig:SkSize_Sims}d-i). In contrast,
m-SK states are consistently absent in $\kappa<1$ dots \textendash{}
both in experiments (\ref{fig:ZFSkyrmions_Expts}b-iii) and simulations
(\ref{fig:SkSize_Sims}d-iii) \textendash{} only single skyrmions
are formed across $W$. These results demonstrate magnetic and geometric
tuning of confined ZF skyrmion stability. 

\begin{comment}
\textbf{\uline{C5. Multiple Skyrmions: Buffer Text}}
\begin{itemize}
\item A
\end{itemize}
\end{comment}

\noindent 
\section{Variation of Skyrmion Size\label{sec:SkSize}}

\noindent %
\begin{comment}
\textbf{\uline{D1. Sk Size Simulations: Buffer Text}}
\begin{itemize}
\item A
\end{itemize}
\end{comment}
\begin{figure*}
\begin{centering}
\includegraphics[width=6.4in]{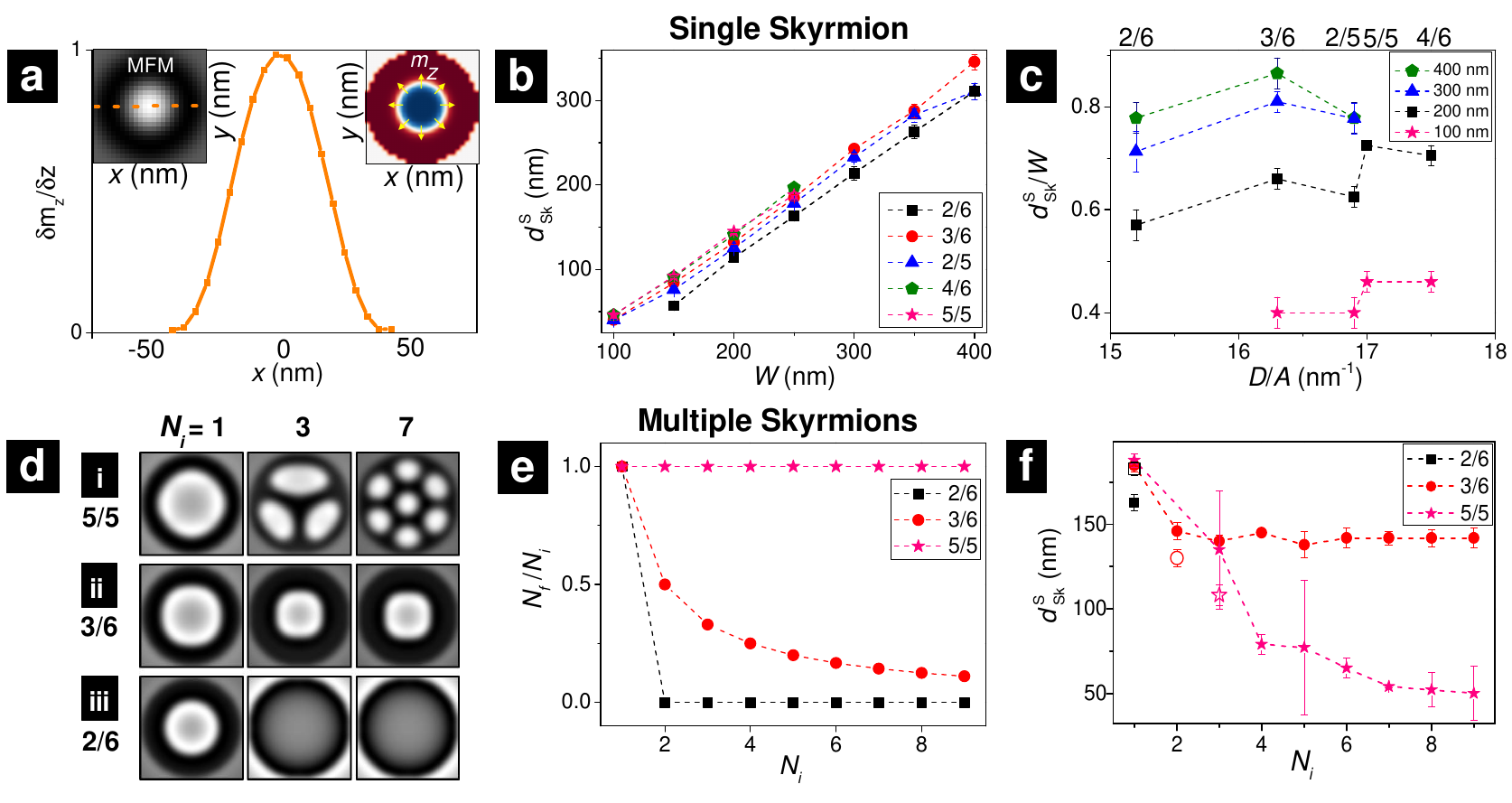}
\par\end{centering}
\noindent \caption[Simulated Variation of Skyrmion Size]{\textbf{Simulated Variation of Skyrmion Size. (a)} Simulated MFM profile
of a Néel skyrmion in a 100~nm Fe(3)/Co(6) dot, extracted from a
linecut (dashed orange) across the simulated MFM image (left inset,
grayscale). A Gaussian fit to the profile gives $d_{{\rm Sk}}^{{\rm S}}$.
Right inset shows the corresponding magnetization ($m_{{\rm z}}$,
color scale), with yellow arrows indicating IP Néel texture. \textbf{(b)}
$d_{{\rm Sk}}^{{\rm S}}$ trend for a single skyrmion initialized
and relaxed in a dot, with varying $W$ across samples. \textbf{(c)}
Modulation of $d_{{\rm Sk}}^{{\rm S}}/W$ (values from b) with $D/A$
across samples. \textbf{(d-f)} Simulated initialization of varying
numbers of skyrmions (number: $N_{{\rm i}}$) in a $W=250$~nm dot
across three samples: Fe(2)/Co(6), Fe(3)/Co(6) and Fe(5)/Co(5). (d)
Grayscale MFM images of the relaxed state for three values of $N_{{\rm i}}$,
showing varying numbers of stabilized skyrmions ($N_{{\rm f}}$).
(e) Observed trend in $N_{{\rm f}}$ for $N_{{\rm i}}$ ranging from
1 to 9. \textbf{(f)} The variation in\textbf{ }$d_{{\rm Sk}}^{{\rm S}}$
with $N_{{\rm i}}$ (solid symbols), with $N_{{\rm i}}>1$ simulations
giving qualitatively different trends across samples. Open symbols
show measured $d_{{\rm Sk}}^{{\rm M}}$ values from experiments for
comparison.\label{fig:SkSize_Sims}}
\end{figure*}

\paragraph{Simulated Single-Sk Sizes}

\noindent A visible modulation in the measured skyrmion size, $d_{{\rm Sk}}^{{\rm M}}$,
across magnetic (vertical) and geometric (horizontal) parameters,
is clearly seen in \ref{fig:ZFSkyrmions_Expts}b. We begin by examining
the trends in simulated skyrmion size, $d_{{\rm Sk}}^{{\rm S}}$ (e.g.
\ref{fig:SkSize_Sims}a, details in \ref{sec:Methods}), to interpret
the experimental trends. \ref{fig:SkSize_Sims}b summarizes the simulated
$d_{{\rm Sk}}^{{\rm S}}$ trends for $N_{{\rm i}}=1$, showing a near-identical
$W$-dependence across samples. This corresponds to a weak dependence
of the normalized size $d_{{\rm Sk}}^{{\rm S}}/W$ on magnetic parameters,
e.g. $D/A$ (\ref{fig:SkSize_Sims}c), consistent with recent multilayer
simulations by other groups\citep{Vidal-Silva2017,Zelent2017}. However,
such insensitivity of $d_{{\rm Sk}}^{{\rm S}}$ to magnetic parameters
is in stark contrast with a visual inspection of MFM data (\ref{fig:ZFSkyrmions_Expts}a-b)
\begin{comment}
\noindent $d_{{\rm Sk}}^{{\rm M}}$ is visibly modulated by magnetic
parameters
\end{comment}
. A reconciliation should be established between simulated and measured
$d_{{\rm Sk}}$ trends by accounting for the aforementioned m-SK stability
for $\kappa\gtrsim1$. %
\begin{comment}
\textbf{\uline{D2. Simulations: Multiple Skyrmions \& Size Evolution:
Buffer Text}}
\begin{itemize}
\item A
\end{itemize}
\end{comment}

\paragraph{Simulated m-SK Sizes}

The relaxed magnetic configuration for $N_{{\rm i}}>1$ simulations
(\ref{fig:SkSize_Sims}d-e for $W=250$~nm) shows a marked transformation
with varying $\kappa$, in line with experimental trends (\ref{fig:ZFSkyrmions_Expts}b).
First, for $\kappa<1$ (\ref{fig:SkSize_Sims}d-iii), only single
skyrmions can be stabilized, and only with $N_{{\rm i}}$$=1$; $N_{{\rm i}}>1$
simulations relax to a UM state. Next, for $\kappa\sim1$ (\ref{fig:SkSize_Sims}d-ii),
$N_{{\rm i}}>1$ simulations relax to a single skyrmion, albeit with
reduced $d_{{\rm Sk}}^{{\rm S}}$. Finally, for $\kappa>1$ (\ref{fig:SkSize_Sims}d-i),
m-SK configurations are formed for $N_{{\rm i}}>1$, while $d_{{\rm Sk}}^{{\rm S}}$
reduces and plateaus for larger $N_{{\rm i}}$. Importantly, these
$d_{{\rm Sk}}^{{\rm S}}$ values (\ref{fig:SkSize_Sims}f: filled)
agree well with measured $d_{{\rm Sk}}^{{\rm M}}$ trends (\ref{fig:SkSize_Sims}f:
empty) when $N_{{\rm i}}$ is appropriately considered. The inclusion
of magnetic granularity\citep{Legrand2017,Juge2017} and interlayer
coupling\citep{Pulecio2016,Zeissler2017} in future could further
improve the quantitative agreement of $d_{{\rm Sk}}^{{\rm S}}$ with
experiments. %
\begin{comment}
\textbf{\uline{D3. Imaging Variations in Sk Size: Buffer Text}}
\begin{itemize}
\item A
\end{itemize}
\end{comment}
\begin{figure*}
\begin{centering}
\includegraphics[width=6.4in]{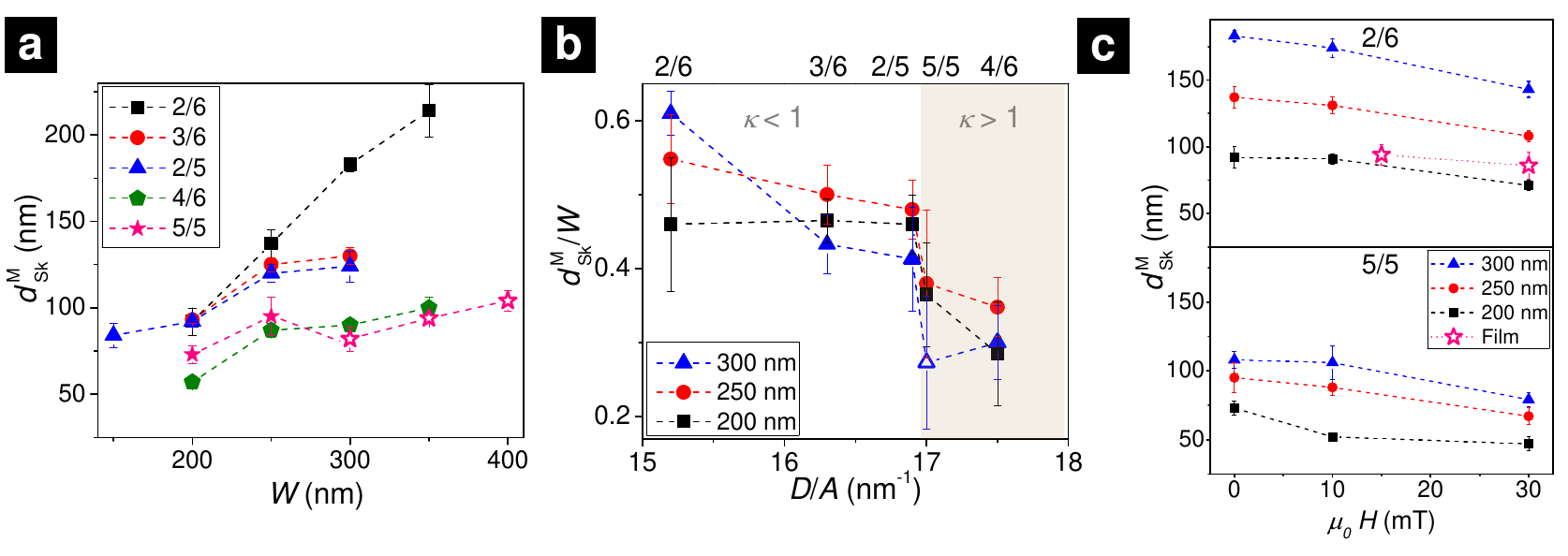}
\par\end{centering}
\noindent \caption[Measured Variation of Skyrmion Size]{\textbf{Measured Variation of Skyrmion Size.} \textbf{(a)} Summary
of measured $d_{{\rm Sk}}^{{\rm M}}$ values at ZF (isotropic Gaussian
fits to MFM data) for all samples across dot sizes. Visible trends
in $d_{{\rm Sk}}^{{\rm M}}$ and extent of $W$-dependence are observed
across samples. Open symbols are derived from m-SK configurations.
\textbf{(b) }The ratio $d_{{\rm Sk}}^{{\rm M}}/W$ for several values
of $W$, plotted against $D/A$ (c.f. \ref{fig:SkSize_Sims}c). \textbf{(c)}
Field dependence of $d_{{\rm Sk}}^{{\rm M}}$ for two representative
samples \textendash{} Fe(2)/Co(6) (film-level results included for
comparison) and Fe(5)/Co(5) \textendash{} for several values of $W$.
\label{fig:SkSize_Expts}}
\end{figure*}

\paragraph{MFM Sk Size Overview}

\ref{fig:SkSize_Expts}a summarizes the measured ZF $d_{{\rm Sk}}^{{\rm M}}$
trends, showing an overall factor of 4 reduction \textendash{} from
$\sim200$~nm down to $\sim50$~nm. Here, the $d_{{\rm Sk}}^{{\rm M}}$
lowest values represent an overestimate from the implicit convolution
with a $\sim30$~nm MFM probe (\ref{sec:Methods}). $d_{{\rm Sk}}^{{\rm M}}$
is monotonically reduced with increased confinement for all samples,
with up to $2.5\times$ reduction for Fe(2)/Co(6). Similarly, for
a given dot size, $d_{{\rm Sk}}^{{\rm M}}$ varies by up to $2.5\times$
across samples. Most interesting in \ref{fig:SkSize_Expts}a is the
marked disparity in the $W$-dependence of $d_{{\rm Sk}}^{{\rm M}}$.
The confinement gradient, defined as $\delta d_{{\rm Sk}}^{{\rm M}}/\delta W$,
reduces by $4\times$ across samples. This observation, while incongruous
with $N_{{\rm i}}=1$ simulations for $\kappa>1$ (\ref{fig:SkSize_Sims}b),
is consistent with the behavior of m-SK states (\ref{fig:SkSize_Sims}f).%
\begin{comment}
\textbf{\uline{D4. Sub-100 nm ZF Skyrmions: Buffer Text}}
\begin{itemize}
\item A
\end{itemize}
\end{comment}

\paragraph{MFM Sk Size Trends}

Finally we examine the evolution of $d_{{\rm Sk}}^{{\rm M}}$ with
$D/A$ (\ref{fig:SkSize_Expts}b) and $H$ (\ref{fig:SkSize_Expts}c),
in the context of extensive predictions of these trends\citep{Sampaio2013,Rohart2013,MoreauLuchaire2016,Mulkers2017,Vidal-Silva2017,Zelent2017,Tomasello2017}.
\ref{fig:SkSize_Expts}b shows that the normalized size, $d_{{\rm Sk}}^{{\rm M}}/W$,
reduces monotonically with increasing $D/A$ \textendash{} with a
sharp jump at $\kappa\sim1$. The sudden factor-of-2 reduction in
$d_{{\rm Sk}}^{{\rm M}}/W$ consistently across all $W$, is indicative
of a fundamental change in the behavior of confined skyrmions around
$\kappa\simeq1$ , especially in light of recent predictions\citep{Zelent2017}.
Next, \ref{fig:SkSize_Expts}c shows the expected reduction in $d_{{\rm Sk}}^{{\rm M}}$
with increasing $H$ across samples. However, the 20-30\% reduction
(for 0\textendash 30~mT) seen here is considerably less than the
$2-3\times$ reduction reported for larger confined skyrmions with
similar fields\citep{Woo2016,Boulle2016,Zeissler2017}. Furthermore,
the ZF trend of $d_{{\rm Sk}}^{{\rm M}}(W)$ is found to persist at
finite fields \textendash{} smaller dots consistently host smaller
skyrmions \textendash{} in contrast with prior reports\citep{MoreauLuchaire2016}.
This highlights the demonstrable robustness of confinement effects
in Ir/Fe/Co/Pt dots. Meanwhile, the marked modulation with magnetic
parameters suggests that a mechanistic understanding of $d_{{\rm Sk}}^{{\rm M}}$,
and its relationship within intrinsic and extrinsic interactions,
merits a detailed theoretical investigation.

\noindent \noindent %
\begin{comment}
\textbf{\uline{D3. Imaging Variations in Sk Size: Buffer Text}}
\end{comment}

\section{Outlook}

\noindent %
\begin{comment}
\textbf{\uline{E1. Summary of Results}}
\begin{itemize}
\item A
\end{itemize}
\end{comment}

\paragraph{Conclusions}

We have presented a comprehensive picture of formation and evolution
of skyrmions at ZF by tailoring confinement effects in Ir/Fe$(x)$/Co$(y)$/Pt
dots. Sub-100~nm skyrmions are stabilized at ZF over a wide range
of magnetic and geometric parameters by intrinsic ($\kappa$), interlayer,
and confinement-induced magnetic interactions. The size of these ZF
skyrmions, here as small as $\sim50$~nm, varies with magnetic and
geometric parameters, by up to $\sim2.5\times$ in either case. Finally,
the ZF stability of multiple skyrmion configurations for $\kappa\gtrsim1$
\textendash{} and the stark contrast in their size evolution across
$\kappa=1$ \textendash{} suggest a strong synergy of thermodynamic
and confinement effects. These results provide a firm foundation for
tailoring the phenomenology of nanoscale skyrmion in confined geometries.
\begin{comment}
\textbf{\uline{E2. Outlook: Skyrmion Physics}}
\begin{itemize}
\item B
\end{itemize}
\end{comment}

\paragraph{Physics Directions}

First, the sub-100~nm Néel skyrmions at ZF reported here show markedly
different physical characteristics from confined skyrmion bubbles\citep{Woo2016,Boulle2016,Zeissler2017}.
This indicates that despite their nominally identical topological
characteristics\citep{Finocchio2016,Fert2017}, exploring the stability\citep{Zelent2017},
detection\citep{Tomasello2017a}, and dynamics\citep{Legrand2017}
of these spin structures could require independent lines of investigation.
Our comprehensive investigation of confined Néel skyrmions offers
a firm foundation tailor-made for such efforts. Second the manifestly
distinct trends in skyrmion configuration and size with varying $\kappa$
go beyond existing predictions\citep{Sampaio2013,Rohart2013,Mulkers2017,Zelent2017,Tomasello2017}.
While recent studies have incorporated the effects of granularity\citep{Juge2017,Zeissler2017},
interlayer coupling\citep{MoreauLuchaire2016}, and dipolar interactions\citep{Zelent2017,Vidal-Silva2017},
we posit that future studies of confined skyrmions would benefit from
harnessing their varying behavior with thermodynamic stability. Finally,
the elastic tuning of skyrmion size with confinement opens up the
exciting possibility of designer magnetic lattices with topological
properties \textendash{} with the potential to engineer frustration,
criticality, and topology under ambient conditions\citep{Nisoli2013}.%
\begin{comment}
\textbf{\uline{E3. Outlook: Skyrmion Technology}}
\begin{itemize}
\item B
\end{itemize}
\end{comment}

\paragraph{Tech Directions}

Crucially, the first realization of sub-100~nm ZF skyrmions in a
device-relevant geometry prompts their immediate employment along
technological lines, especially within perpendicular MTJ devices.
First, the demonstrable modulation of their stability and size with
magnetic interactions and confinement enables mechanistic investigations
of skyrmion creation\citep{Romming2013}, detection\citep{Hanneken2015,Tomasello2017a},
and dynamics\citep{Sampaio2013} in ambient, device-ready conditions.
Next, their sub-100~nm size and ZF stability over a wide range would
enable energy-efficient microwave detectors\citep{Finocchio2015},
oscillators\citep{Garcia-Sanchez2016}, spin valves\citep{Zhou2015},
and magnonic crystals\citep{Chumak2015}. Finally, we note that their
topological stability and malleability with confinement are particularly
suited for highly scalable realizations of random access memory\citep{Apalkov2016}
and synaptic computing\citep{Torrejon2017}. 
\bibliographystyle{apsrev4-1}
\bibliography{SkDots}
\noindent \begin{center}
\rule[0.5ex]{0.6\columnwidth}{0.5pt}
\par\end{center}

\noindent \textsf{\textbf{\small{}Acknowledgments.}}\textsf{\textbf{\emph{\small{}
}}}{\small{}We acknowledge Franck Ernult and Albert Fert for insightful
discussions, and Wen Siang Lew for allowing us to access his instruments.
We acknowledge the support of the A{*}STAR Computational Resource
Center (A{*}CRC), Singapore and the National Supercomputing Centre
(NSCC), Singapore for performing computational work. This work was
supported by the A{*}STAR Pharos Fund (Ref. No. 1527400026) of Singapore,
the Singapore Ministry of Education (MoE), Academic Research Fund
Tier 2 (Ref. No. MOE2014-T2-1-050), and the National Research Foundation
(NRF) of Singapore, NRF \textendash{} Investigatorship (Ref. No.:
NRF-NRFI2015-04). }{\small \par}

\noindent \textsf{\textbf{\small{}Author Contributions.}}\textsf{\textbf{\emph{\small{}
}}}{\small{}P.H., A.S., and C.P. designed and initiated the research.
M.R. deposited the films and characterized them with A.S. and A.K.C.T.
P.H. and L.S.H. fabricated the nanostructures. P.H. and A.K.C.T. performed
the MFM and analyzed the imaging data with A.S. G.S. and A.L.G.O.
performed the micromagnetic simulations. A.S. and C.P. coordinated
the project. All authors discussed the results and provided inputs
to the manuscript.}{\small \par}
\noindent \begin{center}
\rule[0.5ex]{0.6\columnwidth}{0.5pt}
\par\end{center}

\noindent \renewcommand{\thefigure}{M\arabic{figure}}
\renewcommand{\theequation}{M\arabic{equation}}
\renewcommand{\thetable}{M\arabic{table}}
\setcounter{figure}{0}\setcounter{equation}{0}\setcounter{table}{0}

\noindent 
\section{Methods\label{sec:Methods}}

\noindent %
\begin{comment}
\textbf{\uline{E1. Film Deposition \& Characterization}}
\begin{itemize}
\item A
\end{itemize}
\end{comment}
\textsf{\textbf{Film Deposition.}} Multilayer stacks consisting of:
\\
Ta(30)/Pt(100)/{[}Ir(10)/\textbf{Fe($x$)/Co($y$)}/Pt(10){]}$_{20}$/Pt(20)
(nominal layer thicknesses in Å in parentheses) were deposited on
thermally oxidized 100 mm Si wafers by DC magnetron sputtering at
RT using a Chiron\texttrademark{} UHV system manufactured by Bestec
GmbH. The deposition was performed with base pressure below $10^{-8}$~Torr,
and a working pressure of $2\times10^{-3}$~Torr was maintained during
deposition. Five Fe($x$)/Co$(y)$ compositions are investigated here:
Fe(2)/Co(6), Fe(2)/Co(5), Fe(3)/Co(6), Fe(4)/Co(6), and Fe(5)/Co(5).
Varying the Fe($x$)/Co$(y)$ composition enables modulation of the
magnetic parameters \textendash{} $K_{{\rm eff}}$, $D$, and $\kappa$
\textendash{} for this work (see \textcolor{blue}{§SI 1}). These parameters
have been quantified previously\citep{Soumyanarayanan2017}. %
\begin{comment}
\textbf{\uline{F2. Dot Fabrication}}
\begin{itemize}
\item A
\end{itemize}
\end{comment}

\noindent \textsf{\textbf{Dot Fabrication.}} Negative resist Ma-N
2403 was spin-coated on the multilayer films to form a $\sim300$~nm
thick overlayer. Dots of diameter ($W$) 100\textendash 3000~nm were
defined using an Elionix\texttrademark{} electron beam lithography
tool. The patterns were transferred onto the multilayer films using
an Intlvac\texttrademark{} ion beam etching system, with residual
resist lifted off in an ultrasonic bath. Feature topography was imaged
using a Veeco Dimension\texttrademark{} 3100 scanning probe microscope,
and a JEOL\texttrademark{} JSM- 7401 field emission SEM. Cross-sectional
SEM images (e.g. \ref{fig:SkDot_Schematic}c) were obtained by tilting
the sample at nearly 90\textdegree , with the sample mounted on a
vertical holder.\\
Structural characterization of the dots shows an upright profile with
relatively constant diameter vertically through the stacks for $W=200-500$~nm
(see \textcolor{blue}{§SI 2}). For dots with $W\leq150$~nm, a resist
overlayer, due to incomplete lift-off, was observed in numerous cases
(e.g. \ref{fig:ZFSkyrmions_Expts}b-i, $W=150$~nm). The resulting
low yield for $W\leq150$~nm precludes a direct comparison of magnetic
phases across samples.\\
The magnetization properties of $W\geq500$~nm dots were found to
be consistent with film level results (see \textcolor{blue}{§SI 3}).
While AFM images do show some skirting effects at the bottom boundary,
the magnetic layers at the taper are too thin to contribute a detectable
signal in MFM images. %
\begin{comment}
\textbf{\uline{F3: MFM}}

\noindent A
\end{comment}

\noindent \textsf{\textbf{MFM Measurements.}} MFM imaging was performed
using a Veeco Dimension\texttrademark{} 3100 scanning probe microscope,
with Co-alloy coated SSS-MFMR\texttrademark{} tips. The sharp tip
profile (diameter$\sim$30~nm), its ultra-low moment ($\sim$80 emu
cm$^{-3}$), and lift heights of 20-30~nm used during scanning provided
high-resolution MFM images, while introducing minimal stray field
perturbations. The dots were imaged after \emph{ex situ} negative
OP saturation, followed by the application of \emph{in situ} OP fields
ranging from 0 to 200~mT. Repeated MFM scans were acquired to ensure
consistency and reproducibility of results. Twelve dots were imaged
for each $W$ and Fe/Co composition to mitigate variability in deposition
and fabrication processes.%
\begin{comment}
\textbf{\uline{E3. Micromagnetic Simulations}}
\begin{itemize}
\item A
\end{itemize}
\end{comment}

\noindent \textsf{\textbf{Micromagnetic Simulations.}} Micromagnetic
simulations were performed using mumax\textthreesuperior -based simulation
software\citep{Vansteenkiste2014}. The dot was defined with a cylindrical
geometry, in line with experimentally fabricated structures. The mesh
cell size used had lateral dimensions of 2 \textendash{} 4 $\times$
2 \textendash{} 4 nm$^{2}$, while the vertical size $t_{{\rm FM}}$
was set to match the Fe/Co magnetic layer thickness of each sample
(e.g. $t_{{\rm FM}}=0.8$~nm for Fe(2)/Co(6)). A $\sim2$~nm spacer
layer was introduced between magnetic layers (for Ir and Pt), with
the spacer thickness approximated to be the nearest multiple of $t_{{\rm FM}}$.
The results shown correspond to simulations of 20 stack repeats, consistent
with the experimental multilayer film. Single stack simulations were
performed for illustrative comparisons.\\
The magnetic parameters $M_{{\rm S}}$, $K_{{\rm eff}}$, $A$, $D_{{\rm c}}$,
and $D$ used were consistent with our film level results on {[}Ir/Co$(x)$/Fe$(y)$/Pt{]}$_{20}$
stacks\citep{Soumyanarayanan2017} (see \textcolor{blue}{§SI 1}),
and the Gilbert damping parameter $\alpha$ was set to 0.1. A single
(or multiple) skyrmion configuration was initialized in the dot, and
the magnetization was allowed to relax to simulate the ZF configuration
for each set of parameters (see \textcolor{blue}{§SI 4}). For \ref{fig:ZFStates_MuMag}b-c,
bi-linear interpolation was used to map the boundary regions between
the different magnetic states.\\
MFM images were generated from the 2D magnetization profile using
the mumax\textthreesuperior{} built-in MFM function\citep{Vansteenkiste2014},
with a tip height of 20~nm and dipole size of 30~nm respectively.

\begin{figure}
\begin{centering}
\includegraphics[width=2.4in]{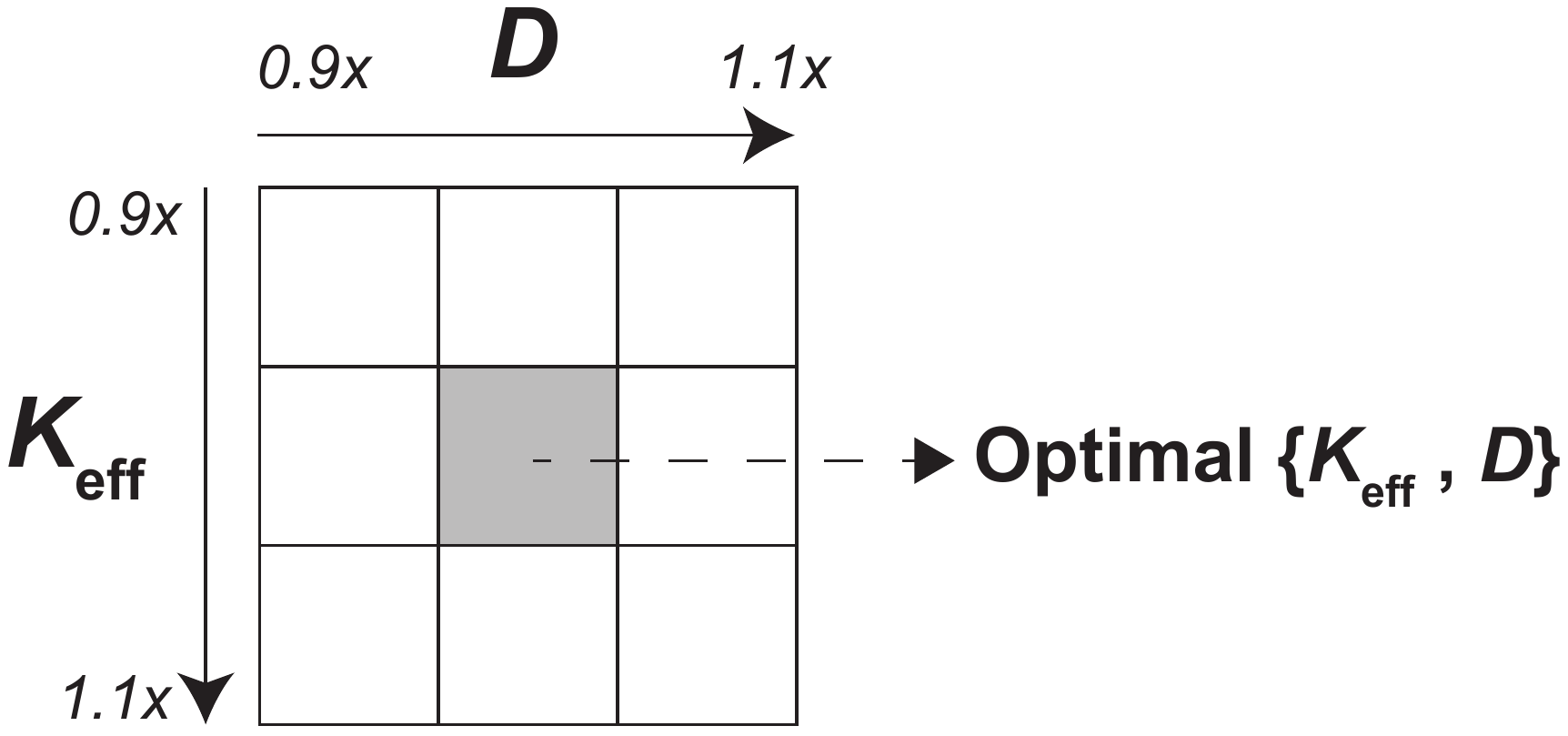}
\par\end{centering}
\noindent \caption[Parametric Array for Phase Diagram]{\textbf{Parametric Array for Phase Diagram. }A weighted average of
a $3\times3$ array, corresponding 10\% variations in $D$ and $K_{{\rm eff}}$
with respect to optimal values (\textcolor{blue}{§SI 1}), is used
to determine the expected magnetic state from micromagnetic simulations
of the dots in \ref{fig:ZFStates_MuMag}d.\label{fig:SkStability_Array}}
\end{figure}

\noindent A weighted average method was adopted to determine the expected
magnetic phase evolution with varying $W$ (\ref{fig:ZFStates_MuMag}d).
A $3\times3$ array of $D$ and $K_{{\rm eff}}$ values was used (\ref{fig:SkStability_Array}),
with the upper and lower limits corresponding to \textpm 10\% with
respect to the optimal values determined previously. The observation
of UM, SK and LS states for each pair of parameters were assigned
weights of -1, 0, and +1 respectively. The total weight of the array
was used to estimate the expected magnetic state for each $W$. 

\end{document}